\documentclass[a4paper]{article}
\normalsize

\begin{document}
\rightline{\small gr-qc/0311075 
\newline \small \hfill FREIBURG-THEP-03-21}
\vspace{0.5cm}
\begin{center}
{\Large\bf Melvin solution with a dilaton potential}
\vspace{0.5cm}

{\bf Eugen Radu}\footnote{E-mail:
{\tt radu@newton.physik.uni-freiburg.de}}
{\bf and Reinoud J. Slagter}\footnote{E-mail:
{\tt rjs@asfyon.nl}}

\vspace*{0.2cm}
{\it $^1$Physikalisches Institut, Albert-Ludwigs-Universit\"at Freiburg} \\
{\it Hermann-Herder-Stra\ss e 3, D-79104 Freiburg, Germany}

{\it $^{2}$University of Amsterdam, Physics Department, The Netherlands}\\
{\it SARA, Stichting Academisch Rekencentrum Amsterdam} and\\
{\it ASFYON, Astronomisch Fysisch Onderzoek Nederland }
\vspace{0.5cm}
\end{center}
\begin{abstract}
We find new Melvin-like solutions in Einstein-Maxwell-dilaton gravity with a
Liouville-type dilaton potential.
The properties of the corresponding solution in Freedman-Schwarz gauged supergravity model
are extensively studied.
We show that this configuration is regular and geodesically
complete but do not preserve any supersymmetry.
An exact solution describing travelling waves in this Melvin-type background
is also presented.
\end{abstract}

\section{ Introduction }

The Melvin magnetic universe is a regular and static, cylindrically symmetric solution
to Einstein-Maxwell theory describing
a bundle of magnetic flux lines in
gravitational-magnetostatic equilibrium \cite{Melvin:1963qx}.
This solution has a number of interesting features,
providing the closest approximation in general relativity for an uniform magnetic field.
The nonsingular  nature of this solution (at the cost of losing the asymptotic
flatness) motivated Melvin to   appoint his solution as a magnetic "geon".
There exist a fairly extensive literature on the properties of
this magnetic universe, including generalizations in several directions.
Rotating and time-dependent magnetic universes have been presented in
\cite{Garfinkle1}, as well as
gravitational waves travelling in a magnetic universe \cite{Garfinkle:dz}.
Of particular interest are black hole solutions in universes which are asymptotically Melvin
\cite{Hiscock:1980zf, Ernst1}.
Recently multidimesional generalizations of the Melvin magnetic universe attracted
much attention
as fluxbranes of the superstring theory.
The Melvin solution has been generalized also
for a gravity theory minimally
coupled to any nonlinear electromagnetic theory, including
Born-Infield theory \cite{Gibbons:2001sx}.

Most of the work on this subject, including pair creation of charged black holes
in a Melvin universe background \cite{Dowker:up,Dowker:bt}
has been done in a Einstein-Maxwell theory and in a generalization
of this theory which include a dilaton $\phi$, whose action is
\begin{eqnarray}
\label{action0}
I= \int  d^4 x \sqrt{-g} \left(\frac{R}{4}-\frac{1}{2}(\nabla \phi)^2
-\frac{1}{4}e^{-2a\phi} F^2 \right),
\end{eqnarray}
where $R$ is the scalar curvature, $F_{\mu \nu}$ is the Maxwell field.
The constant $a$ governs the coupling of $\phi$ to $F_{\mu \nu}$.
For $a=0$ the scalar field decouples and we have the original Einstein-Maxwell theory
with a minimally coupled scalar field,
while $a=1$ is a consistent truncation of the low-energy string theory action.
The value $a=\sqrt{3}$ corresponds to the standard Kaluza-Klein theory.

The dilaton Melvin solution for an arbitrary $a$ takes the form
\cite{Dowker:up,Dowker:bt}
\begin{eqnarray}
\label{MSD}
ds^2=\Lambda^{\frac{2}{1+a^2}}( d\rho^2 + dz^2 - dt^2)
+\Lambda^{-\frac{2}{1+a^2}}\rho^2 d \varphi^2,
\\
\nonumber
e^{-2a(\phi-\phi_0)}=\Lambda^{\frac{2a^2}{1+a^2}},
~~~~~A_{\varphi}=e^{a \phi_0}\frac{B_0 \rho^2}{2\Lambda},
\end{eqnarray}
where
\begin{eqnarray}
\nonumber
\Lambda=1+(\frac{1+a^2}{4})B_0^2 \rho^2.
\end{eqnarray}
The solution is parametrized by $\phi_0$, the value of the scalar field on the symmetry axis and
$B_0$, which characterizes the central strength of the magnetic field.
Although not asymptotically flat, the geometry of this solution is singularity
free and geodesically complete.
A curious property of (\ref{MSD}) is that the total flux
\begin{eqnarray}
\label{flux0}
\Phi=\oint_{\infty} A_{\varphi}=e^{a\phi_0} \frac{4\pi}{1+a^2}\frac{1}{B_0}
\end{eqnarray}
is finite and inversely proportional to $B_0$.
However, in the limit $B_0 \to 0$, even if the geometry becomes flat
and the field strength goes to zero at the centre, the total flux diverges.

A natural generalization of the action (\ref{action0}) is to include a dilaton potential
term $V(\phi)$, which will act such as an effective (position dependent) cosmological
constant.
Black hole solutions for this case have been considered by many authors,
and generally present properties that differ significantly
from the standard Einstein-Maxwell-dilaton theory \cite{Chan:1995fr}.

It is therefore natural to ask whether  magnetic universe solutions also exist
for a nonzero  $V(\phi)$ and
how the properties of the solution are affected by the potential term.
To our best knowledge, to date this question has not been answered in the literature.

It is the purpose of this paper to approach this problem
and to present new Melvin-type exact solution.
Given the inherent difficulties involved in studies of models with nontrivial
dilaton potentials,
the present paper is intended as a first step only and we restrict ourselves
to the case of a Liouville-type potential $V(\phi)=V_0e^{2b\phi}$.
Special attention is paid to the case $a=b=1$ which corresponds to the
the $N=4,~D=4$ gauged $\mathrm{SU(2)}\times\mathrm{SU(2)}$ supergravity, known also as
the Freedman-Schwarz (FS) model \cite{Freedman:1978ra}.
The corresponding Melvin solution has a particularly simple form in this case,
which allows us to discuss its property to some extend.

This paper is organized as follows. In Section 2 we describe
the basic formalism, derive the field equations and present explicit solution in several cases.
In Section 3 we present the Melvin solution in context of the $N=4,~D=4$ gauged
supergravity model
solution and analyse its properties.
We conclude with Section 4 where the results are compiled.

\section{General framework and equations of motion}
The field equations are obtained by varying the action (\ref{action0}) (where we included
the dilaton potential term $V(\phi)$)  with respect
to the field variables $g_{ij}$, $\phi$ and $A_i$
\begin{eqnarray}
\label{eqEinstein}
R_{\mu\nu} - \frac{1}{2}Rg_{\mu\nu} &=&
2\Big(\partial_{\mu} \phi\partial_{\nu}\phi
-\frac{1}{2}g_{\mu\nu}\partial_{\rho}\phi\partial^{\rho}\phi
+V(\phi)g_{\mu\nu}
+e^{-2a \phi} (F_{\mu\rho}F_{\nu}^{~\rho}
- \frac{1}{4}g_{\mu\nu}(F_{\rho\sigma}
F^{\rho\sigma} ) \Big),
\\
\label{eqScalar}
\frac{1}{\sqrt{-g}}\partial_{\mu}(\sqrt{-g}\partial^{\mu}\phi)
&=& -\frac{a}{2} e^{-2a\phi}
F_{\rho\sigma}F^{\rho\sigma}
-\frac{\partial V(\phi)}{\partial \phi},
\\
\label{eqEM}
0 &=& \partial_{\mu}(\sqrt{-g}e^{-2a\phi}F^{\mu\nu}).
\end{eqnarray}
We start by considering a line element on the form
\begin{eqnarray}
\label{metric}
ds^2=M^2(r)dr^2+N^2(r)d\varphi^2+P^2(r)(dz^2-dt^2).
\end{eqnarray}
The symmetries of this line element are the same as that of the original Melvin solution (\ref{MSD}).
There are again four Killing vectors corresponding to translation
along the $t$ and $z$ directions,
a rotation along the $z$ axis and the $t-z$ boost (note that this assures that 
(\ref{metric}) is cylindrically symmetric since it 
admits a $G_2$ on $S_2$ group of isometries
containing an axial symmetry \cite{Carot:1999zm}).
One can fix the residual gauge freedom $r \to \tilde{r}$ of this metric
ansatz  by imposing a gauge condition on the function $M,~N,~P$.

We suppose also that the Maxwell potential presents only on nonvanishing component
$A_{\varphi}(r)$.
Thus the Maxwell equations can easily be integrated to obtain
\begin{eqnarray}
\label{Fik}
F_{r \varphi}=B_0e^{2a \phi}\frac{MN}{P^2},
\end{eqnarray}
where $B_0$ is a constant of integration.
The Einstein and scalar field equations then yields
\begin{eqnarray}
\label{eqs1}
\frac{1}{2} \frac{P'^2}{P^2M^2}+\frac{N'P'}{NPM^2}
&=&\frac{1}{2}B_0^2\frac{e^{2a \phi}}{ P^4}+\frac{\phi'^2}{2M^2}+V(\phi),
\\
\label{eqs2}
\frac{1}{2} \frac{P'^2}{P^2M^2}-\frac{M'P'}{PM^3}+\frac{P''}{M^2P}
&=&
\frac{1}{2}B_0^2\frac{e^{2a \phi}}{ P^4}-\frac{\phi'^2}{2M^2}+V(\phi),
\\
\label{eqs3}
\frac{1}{2} \frac{M'P'}{PM^3}-\frac{1}{2} \frac{P''}{M^2P}
-\frac{1}{2} \frac{N''}{M^2N}-\frac{1}{2} \frac{N'P'}{M^2NP}
+\frac{1}{2} \frac{M'N'}{M^3N}
&=&
\frac{1}{2}B_0^2\frac{e^{2a \phi}}{ P^4}+\frac{\phi'^2}{2M^2}-V(\phi),
\\
\label{eqs4}
\left(\frac{NP^2\phi'}{M}\right)'&=&-aB_0^2e^{2a \phi}\frac{MN}{P^2}-MNP^2\frac{\partial V(\phi)}{\partial \phi},
\end{eqnarray}
where a prime denotes a derivative with respect to $r$.
These equations take a simpler form for the gauge choice
\begin{eqnarray}
M^2=e^{3u+v},~~N^2=e^{u-v},~~P^2=e^{u+v},
\end{eqnarray}
the relation (\ref{Fik}) translating to
\begin{eqnarray}
\label{Fuv}
F_{r \varphi}=B_0e^{2a \phi+u-v}.
\end{eqnarray}
The Einstein and dilaton equations have the simple form
\begin{eqnarray}
\label{eqs}
\nonumber
u''&=&4V(\phi)e^{3u+v},
\\
v''&=&2B_0^2 e^{u-v+2a \phi},
\\
\nonumber
\phi''&=&-aB_0^2e^{u-v+2a\phi}-e^{3u+v}\frac{\partial V(\phi)}{\partial \phi},
\end{eqnarray}
together with the constraint
\begin{eqnarray}
\nonumber
3u'^2+2u'v'-v'^2-4\phi'^2-8e^{3u+v}V(\phi)-4B_0^2e^{u-v+2a\phi}=0.
\end{eqnarray}
\subsection{Exact solutions}
To proceed further, however, we must choose a particular form of $V(\phi)$.
In this paper we specialize to the Liouville potential
$V=V_0 e^{2 b \phi}$, which is simple enough to allow explicit solutions.
This form of potential is also of practical interest, since it frequently arises
in the bosonic sector of gauged supergravities.
Unfortunately,
even for this  simple form of $V(\phi)$,
we were unable to obtain a general solution of the field equations, valid
for every $a,b,V_0$.
However, several special values of these parameters allow for exact solutions.

The scalar field equation  takes a particularly simple form in the $(u,v)$
metric parametrization
\begin{eqnarray}
\phi''=-\frac{1}{2}(av''+bu''),
\end{eqnarray}
and can be integrated one time to $\phi'=-1/2(av'+bu')+const.$
For the choice $const.=0$ the equations (\ref{eqs}) read
\begin{eqnarray}
\label{uv}
u''&=&4V_0e^{2b\phi_0} e^{(3-b^2)u+(1-ab)v},
\nonumber
\\
v''&=&B_0^2e^{2a\phi_0} e^{(1-ab)u-(1+a^2)v},
\end{eqnarray}
with $\phi =-1/2(av +bu )+\phi_0$, $\phi_0$ being an arbitrary constant.
The system (\ref{uv}) admits exact solutions for special values of $a, b$.

\subsubsection{Kaluza-Klein solutions}
In the case $a=1/b=\sqrt{3}$ the action of the theory
corresponds to the Kaluza-Klein reduction
of the five dimensional gravity with a cosmological constant
\begin{eqnarray}
\label{action5}
I= \int d^5 x \sqrt{-g_5}\frac{1}{4}\left(R_5-2\Lambda_5\right),
\end{eqnarray}
in which the five-metric is parametrized as follows
\begin{eqnarray}
\label{metric5d}
ds_5^2=e^{-\frac{4\phi}{\sqrt{3}}}(dx^5+2A_{\mu}dx^{\mu})+
e^{\frac{2\phi}{\sqrt{3}}}g_{ij}dx^i dx^j.
\end{eqnarray}
The value of the cosmological constant fixed the value $V_0$ through
$V_0=- 2\Lambda_5$.
Therefore one may hope to generate Melvin-like dilaton solutions
starting with suitable vacuum (anti-)de Sitter five dimensional configurations.

For this choice of $(a,b)$, an exact solution of the equations (\ref{uv}) reads
\begin{eqnarray}
\label{soluv}
u=u_0-\frac{3}{4}\log \sinh (r+c_1),
~~~
v=v_0+\frac{1}{2}\log \cosh (r+c_2),
~~~
\phi=\phi_0-\frac{\sqrt{3}}{2}(v+\frac{u}{3}),
\end{eqnarray}
where $\phi_0,~c_1,~c_2$ are arbitrary constants,
$e^{-4v_0}=2B_0^2e^{2\sqrt{3}\phi_0},~~e^{8u_0/3}=3e^{-2\phi_0/\sqrt{3}}/(16V_0)$,
while the magnetic field is given by (\ref{Fuv}).

However, different from the $\Lambda_5 \to 0$ limit, the solution in this case present
some unphysical properties
and seems to be less physical relevant.
In particular, similar to other cases \cite{mars}, this solution presents 
a singular axis of symmetry, 
since  the limit 
$\lim_{r \to  r_0} g_{\varphi \varphi}/((r-r_0)^2g_{r r})$ (where $r_0$ is a zero of
the metric function $g_{\varphi \varphi}$)
is singular for every choice of $c_1,~c_2$.
Therefore, this is not the type of configuration we are interested.

\subsubsection{The case $a=b$}
The particular case $a=b$ allows  us also to find
an exact solution.
The gauge choice $M=1/N$ implies from the Einstein equations
 (\ref{eqs1}) and (\ref{eqs2}) the simple relation
$ P''/P =-\phi'^2$,
We further assume  $P(r)=r^n$, which leads
to the solution
\begin{eqnarray}
\label{NMP}
N^2(r)&=&\frac{1}{M^2(r)}~=~Cr^{1-2n}
+\frac{2V_0e^{2a\phi_0}}{n(4n-1)}r^{2n}
-B_0^2e^{2 a\phi_0}\frac{r^{-2n}}{n},~~~P(r)=r^n,
\\
\nonumber
\phi&=&\phi_0\pm \sqrt{n(1-n)}\log r,~~
F_{r\varphi}=B_0 e^{2a\phi}r^{-2n},
\end{eqnarray}
$C$ is an arbitrary constant of integration.
The Einstein equations impose also
\begin{eqnarray}
a=b=\mp \sqrt{(1-n)/n},
\end{eqnarray}
which implies that we must restrict to $0<n<1$ and $a\neq \sqrt{3}$.

The parameters $V_0,~C$ in the expression of $N^2$ should be restricted such that
$N^2(r)>0$ asymptotically.
For $a^2>3$, the term $r^{1-2n}$ is dominant for large values of $r$ and we should impose
that $C>0$.
For values of $a^2<3$, the second term  will be dominant as $r \to \infty$ and the
value of $V_0$ should be positive.
We can also proven that this solution presents a singularity as $r \to 0$ and
the function $N(r)$ has at least one zero.
Therefore the $g_{rr}$ metric function is negative  for $r<r_0$ and positive for $r>r_0$
($r_0$ should be considered the first solution
of the equation $N(r)=0$ when coming from infinity).
However, when $g_{rr}$ becomes negative so does $g_{\varphi \varphi}$
and this  leads  to an apparent change of
signature from $+2$ to $-2$.
This indicates that,  comparable   to similar situations (see $e.g.$ \cite{Dias:2002ps}),
an incorrect extension is being used and one should
choose a different continuation to describe the region $r <r_0$.
Therefore the physical region of our solution is $r\geq r_0$.
The singularity at $r=r_0$ generally corresponds to the origin of the flat space in
polar coordinates.
In the neighbourhood of $r=r_0$ we introduce a new coordinate
$\rho$ defined by
\begin{eqnarray}
\label{rho}
\rho=2 \sqrt{\frac{r-r_0}{g'_{\varphi \varphi }(r_0) } },
\end{eqnarray}
and the $r-\varphi$ section of the metric (\ref{metric}) reads
\begin{eqnarray}
\label{rho-m}
d\sigma_1^2\simeq d\rho^2+
\left(g'_{\varphi \varphi }(r_0) \right)
\rho^2 d\varphi^2.
\end{eqnarray}
Hence the regularity condition as $r \to r_0$ imposes to identify the
 coordinate $\varphi$ with a period
$4\pi r_0 ne^{-2 a \phi_0}/(B_0^2r_0^{-2n}+2V_0r_0^{2n})$.
As $r \to r_0$ the $z-t$ sector of the metric (\ref{metric}) is also regular and
reads $d\sigma_2^2\simeq r_0^2(dz^2-dt^2)$.

Here we note that, if one insists that the coordinate
$\varphi$ is identified with
a period $2 \pi$, this metric has the property of a cosmic string as is 
found in the Abelian Higgs
model (see $e.g.$ \cite{Slagter:1995}).
There will appear an angle deficit (flat space minus a wedge),
if $ r_0 ne^{-2 a \phi_0}/(B_0^2r_0^{-2n}+2V_0r_0^{2n})\equiv G\mu <\frac{1}{2}$, where
$\mu$ is the linear mass density of the string and $G$ the gravitation constant.
In general, it would  be of interest to compare this model with the Higgs model, where the
magnetic flux (see eqs. (\ref{flux0}) and (\ref{fFS})) 
is quantized and depends on the number of zero's of the Higgs field.

However, the general properties of these solutions are rather difficult to discuss for a generic $a$
(in particular we cannot write a form of the line element similar to (\ref{MSD})).
Therefore we prefer to focus for the rest of this paper
on the particular case $a=b=1$.

We close this Section by remarking that the solution (\ref{NMP})
and a class of topological black hole solutions
in Einstein-Maxwell theory with a Liouville-type dilaton potential
\cite{Cai:1997ii}
shares the same Euclidean section and are related by an analytical continuation.
The double "Wick rotation"  $t \to i y,~\varphi \to i\tau$ , together with
$B_0 \to iQ$ in  (\ref{metric})
leads to the solution
\begin{eqnarray}
\label{bh1}
\nonumber
ds^2=\frac{dr^2}{F(r)}+r^n(dy^2+dz^2)-F(r)d\tau^2,
\\
\phi=\phi_0\pm \sqrt{n(1-n)}\log r,~~
F_{r\tau}=Q e^{2a\phi}r^{-2n},
\\
\nonumber
F(r)=~Cr^{1-2n}
+\frac{2V_0e^{2a\phi_0}}{n(4n-1)}r^{2n}
+Q^2e^{2 a\phi_0}\frac{r^{-2n}}{n},
\end{eqnarray}
(where again $a=b=\mp \sqrt{(1-n)/n}$),
corresponding to a topological black hole configuration,
 whose properties are discussed in \cite{Cai:1997ii}.
Here it may be useful to remark that the all known exact solutions
of the  Einstein-Maxwell-dilaton theory with a dilaton potential 
have been obtained within a restricted ansatz.
Different from the asymptotically flat case \cite{Mars:2001pz}, 
the question of solution's uniqueness  
is not answered in the presence of a dilaton potential.
 Thus, other black hole solutions are  likely to exist apart from 
(\ref{bh1}), which, after a a suitable analytical continuation, will lead to
new Melvin-type configurations. 

\section{Melvin solution in Freedman-Schwarz model }
For $a=b=1$, the action (\ref{action})
(with the gauge potential term $V(\phi)=V_0e^{2\phi}$ included)
corresponds to a consistent truncation of the  bosonic sector
of the $N=4,~D=4$ gauged $\mathrm{SU(2)}\times\mathrm{SU(2)}$ supergravity
known also as the Freedman-Schwarz model \cite{Freedman:1978ra}.

The action of the FS model
includes a vierbein $e_{\mu }^{m}$, four Majorana spin-3/2 fields
$\psi _{\mu }^{\rm{I}}$, vector
and six vector fields $A_{\mu }^{a}$ and $B_{\mu }^{a}$ ($a=1,2,2$)
with independent gauge coupling constants $g_{1}$ and $g_{2}$, respectively,
four Majorana spin-1/2 fields $\chi ^{\rm{I}}$, the axion $\eta$
and the dilaton $\phi$ \cite{Freedman:1978ra}.
The bosonic part of the action reads
\begin{eqnarray}
\label{action}
S=\int d^4x   \sqrt{-g}  \Big (\frac{1}{4} R
-\frac{1}{2}(\partial_\mu\phi \,\partial^\mu\phi
+{\rm e}^{4\phi}\partial_\mu\eta \,\partial^\mu\eta)
-\frac{1}{4}\,{\rm e}^{-2\phi}(F^a _{\mu\nu} F^{a \mu\nu}+G^a _{\mu\nu} G^{a \mu\nu})
+\frac{\lambda^2}{8}{\rm e}^{2\phi} \Big),
\end{eqnarray}
where
$
F^a_{\mu\nu} = \partial_{\mu}A^a_{\nu} - \partial_{\nu}A^a_{\mu}
                 + g_1 \epsilon_{abc}A^b_{\mu}A^c_{\nu},
~G^a_{\mu\nu} = \partial_{\mu}B^a_{\nu} - \partial_{\nu}B^a_{\mu}
                 + g_2 \epsilon_{abc}B^b_{\mu}B^c_{\nu}.
$
The gauge coupling constants of the theory fix the value of $V_0=\lambda^2/8$
(with $\lambda^2=g_1^2+g_2^2$, thus $V(\phi)$ corresponding to a negative effective cosmological
constant).

A review of the known exact solutions of this theory
should start with  stable electrovac configuration
found by Freedman and Gibbons \cite{Freedman:xa}, which is a product manifold
$AdS_2\times R^2$, and preserves one quarter or one half
of the supersymmetries, the latter case occurring if one of the two
gauge coupling constants vanishes.
There are also other supersymmetric vacua of the FS model,
in particular the domain wall solution \cite{Cowdall:1998bu,Singh:1998qd}
preserving also one half of the supersymmetries.
This solution has vanishing gauge fields and is purely dilatonic.
Furthermore, BPS configurations involving a non-zero axion were
found by Singh \cite{Singh:1998vf,Singh:1998qd}.
Abelian  BPS black hole solutions  with toroidal event horizon
were constructed in \cite{Klemm:1998in}.
The $N=4,~D=4$ gauged $\mathrm{SU(2)}\times\mathrm{SU(2)}$ supergravity
has also the important  property to allow for exact solutions with
nonabelian matter fields as discovered by Chamseddine and Volkov
\cite{Chamseddine:1998mc,Chamseddine:1997nm} (see also \cite{Radu:2002za}).
It was shown recently that the FS model can be obtained
by compactifying $N=1$ ten dimensional supergravity
on the $\mathrm{SU(2)}\times\mathrm{SU(2)}$ group manifold \cite{ Chamseddine:1998mc,Cowdall:1997fn}
(previously also, a Kaluza-Klein interpretation was given in \cite{Antoniadis:1990mn}).
Therefore it may be worthwhile to look for new exact solutions of FS model,
keeping in mind the recent developments in massive supergravities.

Here we break the gauge group
$\mathrm{SU(2)}\times\mathrm{SU(2)}$ to $\mathrm{U(1)}\times\mathrm{U(1)}$,
considering the Abelian reduction of the
theory (i.e. $A^a_{\mu} = A_{\mu}\delta^{a3},~ B^a_{\mu} = B_{\mu}\delta^{a3}$).
For purely magnetic or electric gauge field ansatz,
$\eta=0$ is a consistent truncation of the FS model.

Since we'll be later interested in the supersymmetric properties of
the Melvin-like solutions,
we present here the general configuration with two electromagnetic fields  
(the one field limit is obvious).
The equations of motion (\ref{eqEinstein})-(\ref{eqEM})
can  easily be generalized to this case,
the only difference being the occurrence
of the corresponding $G_{\mu \nu}$ terms
(the second electromagnetic field  $G_{\mu \nu}$ satisfies an equation similar to (\ref{eqEM})).

In the case $a=b=1$, the field equations  admit the simple solution
\begin{eqnarray}
\label{metricFS}
\nonumber
ds^2&=&\Lambda(d\rho^2+dz^2-dt^2)+
\frac{\sinh^2(c \lambda\rho)}{\lambda^2 c^2 \Lambda } d \varphi^2,
\\
\phi&=&-\frac{\log\Lambda(\rho)}{2}+\log c+\frac{1}{2}\log 2,
\\
\nonumber
A_{\mu}&=&\delta_{\mu\varphi} \frac{2B_0 \cos u}{ \lambda}
\frac{\sinh^2( c \lambda\rho/2)}{\Lambda},
~~~
B_{\mu}=\delta_{\mu \varphi} \frac{2B_0 \sin u}{ \lambda}
\frac{\sinh^2( c \lambda\rho/2)}{\Lambda},
\\
\nonumber
\rm{with~~}\Lambda &=&\frac{\alpha  }{\lambda}
\Big(1+(B_0^2+1)\sinh^2(\frac{c \lambda\rho}{2})\Big),
\end{eqnarray}
where the potential vectors are
given in a regular gauge.
Here $\alpha, u,c$ are arbitrary real constants.
This solution presents a number of similarities with the famous Melvin solution,
which is a solution of $D=4,~N=2$ supergravity.
As expected, in the limit $\lambda \to 0$
(taken together with the rescaling $ B_0 \to B_0\sqrt{2}/(\lambda c)$)
the above solution reduces
to the $a=1$ Melvin solution in (\ref{MSD}).

\subsection{Solution properties}

This spacetime has no horizons and no coordinate singularities except from $\rho=0$.
If we suppose an angular coordinate range $0\leq \varphi<2\pi$ and 
impose the usual regularity condition on the symmetry axis
\begin{eqnarray}
\label{reg}
\lim_{\rho \to  0} \frac{1}{\rho^2} \frac{g_{\varphi \varphi}}{g_{\rho \rho}}=1,
\end{eqnarray}
we find $\alpha=\lambda$.
One can also verify that the ratio $\alpha/\lambda$ has no physical meaning, 
since it can be absorbed in a 
rescaling of the coordinates $x^i$
(for a different periodicity of $\varphi$, a simple rescaling of the angular coordinate
allows (\ref{reg}) to be satisfied)).
Thus, similar to the original dilaton Melvin case,
the regular solution possesses only two essential parameters $c$ and $B_0$,
related to the initial value of the scalar field and the magnitude of the magnetic field.

The flux of the magnetic field
$\Phi=\int (F+G)=\oint_{r \to \infty} (A_{\varphi}+B_{\varphi})$ is also finite
\begin{eqnarray}
\label{fFS}
\Phi=\frac{4\pi B_0 (\cos u+\sin u)}{\lambda(B_0^2+1)},
\end{eqnarray}
and, different from the case without a dilaton potential,
vanishes in the limit $B_0 \to 0$.
This is a manifestation of the different asymptotic structure of spacetime, since
these solutions are not asymptotically flat nor (anti-)de Sitter.

It is easy to check the absence of singularities in scalar polynomials of the
curvature.
We find
\begin{eqnarray}
R=-\frac{c^2 \lambda^2}
{16 \Lambda^3}
 \Big(13-6 B_0^2+13B_0^4
-16(B_0^4-1) \cosh(c\lambda \rho)+3(B_0^2+1) \cosh(2c\lambda \rho)\Big),
\end{eqnarray}
for the curvature scalar, while the explicit expressions for $R_{\mu \nu}R^{\mu \nu}$
and the Kretschmann scalar are very long and will not
be given here.
However, they are both on the form $w(r)/\Lambda^6$, where $w(r)$ is an everywhere finite function.

We can show also that the spacetime described by (\ref{metricFS}) is  both null and
timelike geodesically complete.
For the general line element (\ref{metricFS}),
the equations of
the geodesics have the four straightforward first integrals
\begin{eqnarray}
\label{int1}
\nonumber
P_{z}&=&\Lambda \dot{z},
\\
\nonumber
\label{int2}
L&=&\frac{\sinh^2(c \lambda\rho)}{\lambda^2 c^2 \Lambda }\dot{\varphi},
\\
\label{int3}
E&=&\Lambda \dot{t},
\\
\label{int4}
\nonumber
-\varepsilon&=&\Lambda (\dot{\rho} ^{2} +\dot{z}^{2} - \dot{t}^2)+
\frac{\sinh^2(c \lambda\rho)}{\lambda^2 c^2 \Lambda }\dot{\varphi}^2,
\end{eqnarray}
where a superposed dot stands for as derivative with respect
to the parameter $\tau$ and $\varepsilon=1$ or
$0$ for timelike or null geodesics respectively.
$\tau$ is an affine parameter along the geodesics; for timelike geodesics,
$\tau$ is the proper time.
From the above equations we get the differential equation for the radial coordinate
\begin{eqnarray}
\label{g1}
\left(\frac{d\rho}{d \tau}\right)^2=\frac{E^2-P_z^2}{\Lambda^2}-\frac{\varepsilon}{\Lambda}
-\frac{\lambda^2 c^2 L^2 }{\sinh^2(c \lambda\rho)}.
\end{eqnarray}
Unfortunately, different form the original Melvin solution,
this equation cannot be solved in closed form (except for
$\varepsilon=0$ which leads to a very complicated expression).
However, some general qualitative properties of the geodesic motion
can easily be deduced.
From (\ref{int3}) and (\ref{g1}) we find
\begin{eqnarray}
\label{g2}
\frac{d\rho}{dt}=\Big(
1-\frac{P_z^2}{E^2}-
\frac{\varepsilon \Lambda}{E^2}-\frac{\lambda^2 c^2 L^2}{E^2}
\frac{\Lambda^2}{\sinh^2(c \lambda\rho)} \Big) ^{1/2}.
\end{eqnarray}
It is evident that, since the term $ \Lambda/E^2$ in the above relation diverges asymptotically
while $  \Lambda^2/\sinh^2(c \lambda\rho) $ tends to a constant value in the same limit,
any massive particle  cannot escape to radial infinity 
(a similar similar situation was found in the original Melvin universe \cite{thorne}).
The limit $r \to \infty$ is allowed only for massless particle
satisfying the condition $1- P_z^2/E^2-\lambda^2 c^2 L^2 (B_0^2+1)^2 /(4E^2)>0$.
Also, $\rho=0$ is not an admissible value for particles with nonvanishing angular momentum $L$.

The geodesics of constant $(\rho, z)$ are given by
\begin{eqnarray}
\label{g3}
\varphi=\pm \frac{c \lambda \Lambda}{2 \sinh (c \lambda \rho)}
\Big( \frac{B_0^2+1}{\cosh^4 (c\lambda \rho/2)+B_0^2 \sinh^4 (c\lambda \rho/2) }
\Big) ^{1/2}t,
\end{eqnarray}
and are circles about the symmetry axis.
In this case, different from the Melvin universe,
a free particle can move in a circular orbit
for any finite values of the radial coordinate.

At large $\rho$, the spacetime (\ref{metric}) approaches a conformally flat geometry
\begin{eqnarray}
\label{as}
ds^2\simeq \frac{\alpha (B_0^2+1)}{ 4 \lambda} e^{c \lambda \rho}
\left(d \rho^2+dz^2 
+dw^2
-dt^2
\right),
\end{eqnarray}
where $w=2 \varphi/((B_0^2+1)c \alpha )$ is a compact coordinate.
Thus, our generalized Melvin universe  does not approach one of the dilaton Melvin solutions (\ref{MSD}),
which is clearly a consequence of the presence of a dilaton potential with no fixed points
in the action.

\subsection{A generation procedure}

We can easily prove the following property.
Let $(g_{\mu \nu}, A_\mu, \phi)$ be an axisymmetric
solution of the  field equations (\ref{eqEinstein})-(\ref{eqEM}),
for a Lioville-type potential $V=V_0e^{2b\phi}$
with $b=1/a$.
All the fields are independent
of the azimuthal coordinate $\varphi$.
Let the other three
coordinates be denoted by $\{x^i\}$. Suppose also that
$A_i=g_{i\varphi}=0$.
Then a new solution of the equations of motion is given by
\begin{eqnarray}
\label{transf}
g^\prime_{ij}=\Lambda^{\frac{2}{1+a^2}} g_{ij},~~
g^\prime_{\varphi\varphi}=\Lambda^{-\frac{2}{1+a^2}}g_{\varphi\varphi},~~
e^{-2a\phi^\prime}=e^{-2a\phi}\Lambda^{\frac{2a^2}{1+a^2}},
\\
\nonumber
A^\prime_\varphi=-\frac{2}{(1+a^2)B\Lambda}(1+\frac{(1+a^2)}{2}B A_\varphi),
~~\Lambda=(1+\frac{(1+a^2)}{2}B A_\varphi)^2+\frac{(1+a^2)B^2}{4}g_{\varphi\varphi}e^{2a\phi}.
\end{eqnarray}
As explicitly proven in \cite{Dowker:bt,Llatas:1997pa}, 
the action principle (\ref{action0}) is invariant under the
above transformation, for any values of $a$.
However, we can easily prove that 
$\sqrt{-g'}e^{2\phi'/a}=\sqrt{-g}e^{2\phi/a}$,
and thus
the supplementary Lioville potential term in 
the total action $\int  d^4 x \sqrt{-g} V_0e^{2b\phi}$
preserves also the form for $b=1/a$.

The dilaton Melvin metric (\ref{MSD}) is generated applying this procedure to the Minkowski
spacetime, which is not a vacuum of the theory for $V(\phi) \neq 0$.
However, in our case we can easily generate  magnetic configurations
starting with vacuum dilaton solutions.
For example, the regular Melvin solution of FS model (\ref{metric}) (with $\alpha=\lambda $)
is found by using the  pure dilaton seed  metric $(B_0=0)$
\begin{eqnarray}
\label{metricd}
ds^2&=&\cosh^2(\frac{c \lambda\rho}{2})(d\rho^2+dz^2-dt^2)+
\frac{\sinh^2( c \lambda\rho/2)}{\lambda^2 c^2} d \varphi^2,
\\
\nonumber
\phi&=&- \log (\cosh ( c \lambda\rho/2)) +\log c\sqrt{2}.
\end{eqnarray}
Other magnetic solutions are found by using different seed metrics
known in the literature.

One may hope to generate in this way more complex configurations,
black hole solutions immersed in a magnetic universe being of special interest.
For $\phi=V(\phi)=0$, the seed metric is the Schwarzschild one
and the corresponding solution was constructed about
thirty years ago by Ernst \cite{Ernst1}.

However, all known black hole solutions of the FS model have been found for an
ansatz satisfying $g_{\varphi\varphi}=e^{-2\phi}$.
Therefore the transformation (\ref{transf}) does not lead to new solutions in this case.
The situation is different for $1/b=a\neq 1$, where this generation
procedure leads to nontrivial black hole solutions in a
magnetic universe background.
We hope to come  back on this point in the future.

\subsection{Travelling waves in the magnetic universe}

A travelling wave is a wave that propagates without any change in amplitude or shape,
while the induced perturbation  does not  need to be small.
Similar to the original Melvin universe
and its nonlinear electrodynamics version \cite{Gibbons:2001sx}, the
Melvin-type solution (\ref{metricFS}) allows for travelling waves
generalizations.

To find this solution, we follow the general approach  presented in \cite{Garfinkle1},
and consider a generalized Kerr-Schild metric ansatz
\begin{eqnarray}
\label{Kerr-Schild}
\bar{g}_{\mu \nu}=g_{\mu \nu}+\Lambda^{-2}\Psi(u,\rho,\varphi)k_{\mu}k_{\nu},
\end{eqnarray}
where $g_{\mu \nu},\Lambda$ are those of the background Melvin-type metric (\ref{metricFS}),
$\Psi(u,\rho,\varphi)$ is a scalar whose form is determined by the field equations,
while $k_{\mu}=\delta_{\mu}^u$.
Here we have introduced the null coordinates $(u,v)=(z \pm t)/\sqrt{2}$,
$k_{\mu}$ being a null Killing
vector.

Similar to the cases discussed by Garfinkle and Melvin \cite{Garfinkle1}
or Gibbons and Herdeiro \cite{Gibbons:2001sx},
the field equations are solved if the function
$\Psi(u,\rho,\varphi)$ is harmonic in the unperturbed Melvin-type universe (\ref{metricFS}), $i.e.$
$\nabla^2 \Psi(u,\rho,\varphi)=0$, where $\nabla^2 $ is considered with respect to $g_{\mu \nu}$.
In the new travelling wave solution, the electromagnetic and dilaton field are still given by  (\ref{metricFS}).
By separating variables we  find that
$\Psi(u,\rho,\varphi)=f(u)P(\rho) \cos \nu(\varphi-\varphi_0)$, where
$f(u)$ is an arbitrary smooth function
giving the profile of the wave (for $f(u)=0$ we find the background metric (\ref{metricFS})).

The function $P(\rho)$ is a solution of the equation
\begin{eqnarray}
\label{eq-P}
\sinh(c\lambda \rho)\frac{d}{d \rho}\big(\sinh(c\lambda \rho)\frac{dP(\rho) }{d\rho}\big)
-\frac{c^2\lambda^2\nu^2}{4} \Big(1-B_0^2+(1+B_0^2)\cosh^2(c\lambda \rho)\Big)^2 P (\rho)
=0,
\end{eqnarray}
(a simpler form of this equation is obtained by taking $\cosh(c\lambda \rho)=x$).
The geometry describing travelling waves in a Melvin model
with Lioville dilaton potential is then
\begin{eqnarray}
\label{ge-tw}
d\bar s^2=\Lambda(\rho)\bigg(d\rho^2+2dudv +f(u) \cos \nu(\varphi-\varphi_0)P(\rho)du^2 \bigg)
+\frac{\sinh^2(c \lambda\rho)}{\lambda^2 c^2 \Lambda } d \varphi^2.
\end{eqnarray}
Unfortunately, the equation (\ref{eq-P}) can be solved in closed form for special values of
$(\nu,B_0)$ only.
The solution with $\nu=0$ is
$P(\rho)= c_1\log(\tanh(c\lambda \rho/2))+c_2$, while
for $B_0=1,~\nu=1/2$ we find
$P(\rho)= (c_3 \sinh(c\lambda \rho/\sqrt{2})+c_4 \cosh(c\lambda \rho/\sqrt{2}))/
\sqrt{\sinh(c\lambda \rho)}$, with $c_i$ arbitrary real constants.
We remark that these configurations presents an essential singularity as $\rho \to 0$,
while the second function $P(\rho)$ presents also a minimum for a finite value of $\rho$.
The general solution of the equation (\ref{eq-P}) possesses the same features
as found by Garfinkle and Melvin \cite{Garfinkle1}, implying that for any choice of $\nu$
the metric is singular.

\subsection{Interpretation as solution of $D=10$ supergravity}

As was shown in \cite{Chamseddine:1998mc}, the FS model
can be obtained via dimensional reduction
of the  $N=1,~ D=10$  supergravity, which contains apart from gravity
and the dilaton field an antisymmetric tensor field $\hat{H}_{ABC}$.
As a result, any
on-shell configuration $(g_{\mu\nu},A^{a}_{\mu}, B^{a}_{\mu},\phi)$
in the model (\ref{action}),
can be uplifted to become a solution
of ten-dimensional equations of motion for the $D=10$ supergravity.
The details of the compactification
on the group manifold $S^3 \times S^3$ are given in \cite{Chamseddine:1998mc},
so we shall not repeat them here.
We adopt the index convention used in \cite{Chamseddine:1998mc}, i.~e.~greek and latin
indices refer to the four-dimensional and internal
six-dimensional ($S^3 \times S^3$) spaces, respectively.

In the Einstein frame,
the ten dimensional solution reads
\begin{eqnarray}
\label{metric10d}
ds_{10}^2=e^{-3\phi/2}g_{\mu \nu}dx^{\mu}dx^{\nu}
+2e^{-\phi/2}\left(\Theta^{(1)a}\Theta^{(1)a}+\Theta^{(2)a}\Theta^{(2)a}\right)
,
\end{eqnarray}
where $g_{\mu \nu}dx^{\mu}dx^{\nu}$ is the four dimensional line element, ($a,b,c=1,2,3$),
\begin{eqnarray}
\nonumber
\Theta^{(1)a}= A^{a}+\frac{\epsilon^a}{g_1},
~~~\Theta^{(2)a}= B^{a}+\frac{\epsilon^a}{g_2},
~~~A^{a}=A^{a}_{\mu} dx^{\mu},
~~~B^{a}=B^{a}_{\mu} dx^{\mu},
~~~F^{(1)a}_{\mu \nu}=F^{a}_{\mu \nu},
~~~F^{(2)a}_{\mu \nu}=G^{a}_{\mu \nu},
\end{eqnarray}
 while $\epsilon^a$
are the invariant 1-forms on $S^3$
\begin{eqnarray}
\nonumber
\epsilon^1=\cos\psi d\theta+\sin\psi\sin\theta_1 d\Phi,\
\epsilon^2=-\sin\psi d\theta+\cos\psi\sin\theta_1 d\Phi,\
\epsilon^3=d\psi+\cos\theta d\Phi,
\end{eqnarray}
$\psi,\theta,\Phi$ being the Euler angles on the three sphere.
The $D=10$ dilaton field is $\hat{\phi}=\phi/2$,
while the nonvanishing components of the ten-dimensional antisymmetric tensor field
$H_{ABC}$ are given by
\begin{eqnarray}
H_{\alpha\beta  a}=- \frac{1}{\sqrt{2}} e^{-3\phi/4}F_{\alpha \beta }^a,
~~~~H_{abc}=\frac{1}{\sqrt{2}} e^{3\phi/4} f_{abc},
\end{eqnarray}
where $f_{abc}=f_{abc}^{(s)}=g_s \epsilon_{abc}~(s=1,2)$ are the $\mathrm{SU(2)} $ gauge group structure constants.
Using the rules of \cite{Chamseddine:1998mc},
one can further lift the solutions to eleven dimensions
to regard them in the context of M-theory.

\subsection{On the supersymmetry of Melvin solution in FS model}

We now want to study if the solution (\ref{metricFS})   preserve any
amount of supersymmetry.
For the rest of this section we use conventions
similar to \cite{Klemm:1998in} and \cite{Chamseddine:1997nm}, in particular
a mostly minus signature signature
($i.e.$ we take $g_{\mu \nu} \to -g_{\mu \nu}$;
the scalar and electromagnetic field preserve the form given above).
We use   $\{\gamma_{\mu},\gamma_{\nu}\} = 2g_{\mu\nu}$,
$\sigma_{\alpha\beta} = \frac{1}{4}[\gamma_{\alpha},\gamma_{\beta}]$, $\gamma_5 = i\gamma^0\gamma^1
\gamma^2\gamma^3$, so that $\gamma_5^2 = 1$.

One of the supersymmetry transformations for a purely bosonic background and a vanishing axion read
\begin{eqnarray}
\label{transf1}
\delta\bar{\chi} = \bar{\epsilon}
\Big (\frac{i}{\sqrt 2}
(\partial_{\mu}\phi) \gamma^{\mu} -
                     \frac{1}{2}e^{-\phi}
(\alpha^a F^a_{\mu\nu} + i\gamma_5\beta^a G^a_{\mu\nu})
                     \sigma^{\mu\nu} + \frac{1}{4}e^{\phi}
                     (g_1 + i\gamma_5g_2)
\Big ),
\end{eqnarray}
where $\epsilon = \epsilon^I$ are four Majorana spinors.
Here $\alpha^a$ and $\beta^a$ denote the generators of
the (1/2,1/2) representation of
$\mathrm{SU(2)}\times\mathrm{SU(2)}$. We use the form of these matrices
given in \cite{Freedman:1978ra}, in particular
\begin{eqnarray}
\displaystyle
\alpha^3 = \left(\begin{array}{cc}
                      i\sigma_2 & 0 \\
                      0 & i\sigma_2
                      \end{array} \right),\qquad
\beta^3 = \left(\begin{array}{cc}
                      i\sigma_2 & 0 \\
                      0 & -i\sigma_2
                      \end{array} \right).
\end{eqnarray}
The condition that the variation of the Majorana field $\chi$ be vanishing can be written as
\begin{eqnarray}
\bar{\epsilon}({\cal M}_1 + \alpha^3{\cal M}_2 + \beta^3{\cal M}_3) = 0,
\label{fermvar}
\end{eqnarray}
where the ${\cal M}_i$ are $4\times 4$ matrices given by
\begin{eqnarray}
{\cal M}_1 &=& \frac{i\Lambda'}{(2\Lambda)^{3/2}}\gamma_1 + \frac{1}{4}
               \sqrt{\frac{2c^2}{\Lambda}}(g_1 + i\gamma_5 g_2),
\nonumber
\\
{\cal M}_2 &=& -\frac{cB_0\alpha \cos u }{(2\Lambda)^{3/2}}\gamma_1 \gamma_3,
\\
{\cal M}_3 &=& \frac{cB_0\alpha \sin u }{(2\Lambda)^{3/2}}i \gamma_5 \gamma_1 \gamma_3.
\nonumber
\end{eqnarray}
which implies
\begin{eqnarray}
\nonumber
\bar{\epsilon}_{^{}}\Theta = 0,
\end{eqnarray}
the $16\times 16$ matrix $\Theta$ being defined as
\begin{eqnarray}
\nonumber
\Theta = \left(\begin{array}{cc}
                      \Theta_+ & 0 \\
                      0 & \Theta_-
                      \end{array} \right), \qquad
\Theta_{\pm} = \left(\begin{array}{cc}
                      {\cal M}_1 & {\cal M}_2 \pm {\cal M}_3 \\
                      -({\cal M}_2 \pm {\cal M}_3) & {\cal M}_1
                      \end{array} \right).
\end{eqnarray}
The necessary condition for the existence of Killing spinors $\bar{\epsilon}$
is thus $\det\Theta = 0$
which yields
\begin{eqnarray}
\label{cond}
B_0^2(g_2 \cos u\pm g_1 \sin u)^2+\frac{\lambda^2}{4}(B_0^2-1)^2=0.
\end{eqnarray}
Therefore the conditions   $g_2 \cos u=\pm g_1 \sin u$ and $B_0^2=1$
should be satisfied by the supersymmetric
solutions.
However, we prove that these conditions are too strong in order to allow for
Killing spinors to exist.
Returning to the equation (\ref{fermvar}), we  write it in the form
\begin{eqnarray}
\label{new}
\frac{\Lambda'}{c}+\alpha B_0 \hat{O_1}+\lambda \Lambda \hat{O_2}=0
\end{eqnarray}
where
\begin{eqnarray}
\hat{O_1}= i \alpha^3 (\cos u -i\gamma_5 \sin u)\gamma_3,~~~
\hat{O_2}=\frac{i}{\lambda}(g_1+i\gamma_5 g_2)\gamma_1
\end{eqnarray}
satisfy the obvious relations
$\hat{O_1}^2=\hat{O_2}^2=I,~\{\hat{O_1},\hat{O_2}\}=-2(g_1\cos u-g_2\sin u)/ \lambda I$ (where $I$ is the
$4\times 4$ unit matrix.
One can easily show that  (\ref{new}) implies
\begin{eqnarray}
\label{new1}
\frac{\Lambda'^2}{c^2}=(B_0\alpha \cos u-g_1\Lambda)^2+(B_0\alpha \sin u+g_2\Lambda)^2
\end{eqnarray}
which is not compatible with the constraint (\ref{cond}).
We conclude that, similar to the $a=0$ Melvin solution \cite{Gibbons:2001sx}, 
the configuration
(\ref{metricFS}) does not preserve any supersymmetry.

\section{Conclusions}

Among configurations of fields that are in static equilibrium under their own gravitational
attraction one of the simplest is a parallel bundle of magnetic flux.
The corresponding configuration, known as Melvin solution, have found recently considerable
attention in string/M theory context.

In this paper we have generalized the Melvin solution
by including a Liouville-type dilaton potential in the action principle.
Although it was not possible to solve the field equations in the general case,
we have presented explicit solutions for several values of the
theory constants.
The Melvin solution in $D=4,~N=4$ gauged supergravity has been extensively studied,
and have many similar properties with the known solutions without a dilaton potential.
In particular, we haven't found any fraction of supersymmetry
being left unbroken for this configuration.

We did not considered 
the stability of our solutions against small radial, time-dependent perturbations.
However, since the pure Melvin solution (which is recovered in the limit of vanishing
dilaton potential)
was shown to be stable \cite{thorne,mpr}, one may expect that
the FS configuration (\ref{metricFS}) to be also stable.
However, it could decay by instanton processes similar to the case of vanishing
dilaton potential.
\\
\\
{\bf Acknowledgement}
\newline
 This work of E. Radu was performed in the context of the
Graduiertenkolleg of the Deutsche Forschungsgemeinschaft (DFG):
Nichtlineare Differentialgleichungen: Modellierung,Theorie, Numerik, Visualisierung.
\\

\end{document}